\documentclass[preprint, 12pt]{elsarticle}
\pdfoutput=1
\usepackage{gensymb, footnote}
\usepackage{caption}
\usepackage{subcaption}
\usepackage{amsmath}    % need for subequations
\usepackage{graphicx}   % need for figures
\usepackage{verbatim}   % useful for program listings
\usepackage{color,float}% use if color is used in text
\usepackage{comment}
\journal{Scripta Materialia}
\begin{document}
\begin{frontmatter}

\title{Synchrotron x-ray diffraction studies of the $\alpha \rightleftharpoons \beta$ structural phase transition in Sn and Sn-Cu}
\author[ino,hbni]{A.~Mazumdar\corref{cor2}}
\author[dcmp]{A.~Thamizhavel\corref{cor2}}
\author[dnap]{V.~Nanal\corref{cor1}}
\cortext[cor1]{Corresponding author}
\ead{nanal@tifr.res.in}
\author[iit]{R.G.~Pillay\corref{cor2}}
\author[rrcat]{A.~Upadhyay\corref{cor2}}
\author[ino,hbni]{V.~Vatsa\corref{cor2}}
\author[dnap]{A.~Reza\corref{cor2}}
\author[npd,hbni]{A.~Shrivastava\corref{cor2}}
\author[dcmp]{Bhagyashree~Chalke\corref{cor2}}
\author[dnap]{S.~Mallikarjunachary\corref{cor2}}

\address[ino]
{India-based Neutrino Observatory, Tata Institute of Fundamental Research, Mumbai~-~400005, India}
\address[hbni]
{Homi Bhabha National Institute, Anushaktinagar, Mumbai - 400094, India}
\address[dcmp]
{Department of Condensed Matter Physics and Material Science, Tata Institute of Fundamental Research, Mumbai - 400005, India}
\address[dnap]
{Department of Nuclear and Atomic Physics, Tata Institute of Fundamental Research, Mumbai - 400005, India}
\address[iit]
{Department of Physics, Indian Institute of Technology Ropar, Rupnagar - 140001, India}
\address[rrcat]
{Synchrotron utilization section, Raja Ramanna Centre for Advanced Technology, Indore~-~452013, India}
\address[npd]
{Nuclear Physics Division, Bhabha Atomic Research Centre, Mumbai - 400085, India}

\begin{abstract}
The transformation between the metallic ($\beta$) and semi-conducting ($\alpha$) allotropes of tin is still not well understood. The phase transition temperature stated in the literature, 286.2~K, seems to be inconsistent with recent calorimetric measurements. In this paper, this intriguing aspect has been explored in Sn and Sn-Cu (alloyed 0.5 \% Cu by weight) using temperature resolved synchrotron x-ray diffraction measurements performed at the Indus-2 facility. Additionally, the $\alpha \rightleftharpoons \beta$ Sn transition has been recorded using in-situ heating/cooling experiments in a scanning electron microscope. Based on these measurements, a protocol has been suggested to reduce the formation of $\alpha$-Sn in potentially susceptible systems. This will be useful in experiments like \textit{TIN.TIN} (The INdia-based TIN detector), which proposes to employ $\sim$100 - 1000~kg of superconducting tin-based detectors to search for neutrinoless double beta decay in the isotope $\mathrm{^{124}Sn}$.
\end{abstract}

\begin{keyword}
tin pest \sep polymorphic phase transformation \sep scanning electron microscopy (SEM) \sep synchrotron radiation \sep X-ray diffraction (XRD)
%Need to find the fifth word. This could be 
\end{keyword}

\end{frontmatter}

While tin is commonly regarded as the first metal of Group~14 of the periodic table, a semiconducting allotrope of tin ($\alpha$-Sn) also exists. $\alpha$-Sn has generated a lot of interest recently due to its tunable topological properties and is the only known elemental member of a class of 3-D topological materials called Topological Dirac Semimetals~\cite{TDS_PRL,TDS_APL,TDS_ACS}. The structural phase transformation from the metallic $\beta$-Sn to the semiconducting $\alpha$-Sn is known as tin pest~\cite{TPRev} and this transformation occurs very close to room temperature in ambient conditions. During this transformation, the crystal structure of tin changes from body-centred tetragonal to diamond cubic and it is associated with a 27\% increase in the volume of the unit cell. The brittle semi-conducting $\alpha$-Sn cracks due this sudden volume expansion. Thus, tin pest affects the structural integrity of tin and is a concern in tin-based systems that operate at low temperatures for a long time~\cite{TPRev,TPFailure,TPMRX}. This is also of concern for low temperature electronic circuits that use lead-free solders~\cite{TP2007,TP2008,TP2009,TPSnCu}.

The behaviour of the $\alpha \rightarrow \beta$ (heating) and the $\beta \rightarrow \alpha$ (cooling) processes are different, due to differences in the underlying mechanisms~\cite{TPMech}. Notably, the  $\beta \rightarrow \alpha$ transition starts with a nucleation step. This nucleation can occur spontaneously or can be induced with an isomorphic external seeding agent such as $\alpha$-Sn, InSb or CdTe. The time taken for spontaneous nucleation varies immensely and could take anywhere from a few hours to a few years. If a suitable nucleant is absent, tin may remain in the $\beta$-phase even after cooling past the transition temperature.

\textit{TIN.TIN} (The INdia-based TIN detector)~\cite{TINTIN} proposes to use superconducting tin-based calorimetric detectors to search for neutrinoless double beta decay in $\mathrm{^{124}Sn}$. Since this is a very rare process, the total detector mass is expected to be large ($\sim$100 - 1000~kg) and such experiments also run for several years. During each thermal cycling from room temperature to sub-kelvin temperatures and vice versa, the detector will be susceptible to tin pest. Thus, to maintain the longevity and performance of the detector, it is crucial that the risk of $\alpha$-Sn formation is suitably minimized.

Although 286.2~K is often quoted in the literature as the $\alpha \rightleftharpoons \beta$ Sn phase transition temperature, differential scanning calorimetry (DSC) measurements indicate a higher phase transition temperature. It should be mentioned that this quoted value is based on very old dilatometric measurements~\cite{TPDilatometry}. The peak temperatures of the $\alpha \rightarrow \beta$ Sn phase transition reported by various DSC measurements are summarized in Table \ref{tab:tab1}.
%%%%%%%%%%%%%%%%%%%%%%%%%TABLE1 STARTS HERE%%%%%%%%%%%%%%%%%%%%%%
\begin{table}[h]
\centering
\caption{\label{tab:tab1}The peak temperatures of the $\alpha \rightarrow \beta$ Sn transition measured by various differential scanning calorimetry (DSC) experiments.}
\begin{tabular}{cc}
\hline 
Reference & Peak temperature \\ 
\hline 
Zuo $\textit{et. al.}$~\cite{P20} & 301~K \\
Ojima $\textit{et. al.}$~\cite{P12} & 308 - 313~K \\  
Mazumdar $\textit{et. al.}$~\cite{TPMRX} & 312~K \\ 
Zeng $\textit{et. al.}$~\cite{P29} & 315 - 320~K \\ 
Gialanella $\textit{et. al.}$~\cite{P13} & 338~K \\ 
\hline
\end{tabular}
\end{table} 
%%%%%%%%%%%%%%%%%%%%%%%%%TABLE1 ENDS HERE%%%%%%%%%%%%%%%%%%%%%%
In the previous DSC study~\cite{TPMRX}, an onset phase transition temperature of 307.4~K was reported for Sn, which was consistent with the observations of Ojima $\textit{et. al}$~\cite{P12}. The phase transition temperature for Sn-Cu (alloyed 0.5~\% Cu by weight) was also measured to be 309.3~K. The differences between the measured transition temperatures for Sn and Sn-Cu were within experimental errors, implying that alloying with 0.5~\% Cu by weight does not have any effect on the phase transition temperature.

Since the static lattice energy difference between $\alpha$-Sn and $\beta$-Sn is estimated to be only $\sim$ 10 - 40~meV/atom, density functional theory calculations for this transition are challenging. This calculation has been rigorously pursued and several calculations exist in the literature~\cite{P1, P5, P7}. The typical calculated phase transition temperatures are not consistent with measurements. An exception is the calculated transition temperature of $\sim$ 311~K by Pavone $\textit{et. al.}$~\cite{P7}., which is similar to that measured by Ojima $\textit{et. al.}$~\cite{P12}, Mazumdar $\textit{et. al.}$~\cite{TPMRX} and Zeng $\textit{et. al.}$~\cite{P29}. It has been recently pointed out that this phase transition provides a sensitive test of the accuracy of density functionals and computational methods~\cite{P1}. Therefore, a precision measurement of $\alpha \rightleftharpoons \beta$ transition in tin is highly important.

While some earlier works use x-ray diffraction (XRD) to study the kinetics of the transformation~\cite{P29, P9}, no studies on the phase transition temperature using this technique exist. In this work, the phase transition in Sn and Sn-Cu has been studied using x-ray diffraction (XRD) and scanning electron microscopy (SEM) techniques. The results have been compared with the previously reported DSC measurements in~\cite{TPMRX}.

Sn and Sn-Cu (0.5\% Cu by weight) samples were prepared in the $\alpha$-phase by incubating the samples at a temperature $T\leq 253~K$, adding a small mass of $\alpha$-Sn powder as a seed. The details of the sample fabrication can be found in~\cite{TPMRX}. Grinding the samples for the XRD measurements was infeasible, as this process inevitably caused some of the sample to reconvert to the $\beta$-phase. Instead, the natural disintegration caused by the transformation process from $\beta \rightarrow \alpha$ phase was relied upon for the formation of a granular powder.

Since the volume change associated with the phase transition in tin is significant, the transition can be studied by measuring the volume changes in a tin sample. It should be noted that the quoted value in the literature (i.e., 286.2~K) is based on dilatometry experiments~\cite{TPDilatometry}, in which the tin samples were placed in a fluid medium and the expansion/contraction of the sample was recorded by measuring the change in the height of the liquid meniscus. Analogous to dilatometric experiments, in this paper, in-situ temperature-resolved SEM images of tin samples were recorded in order to observe the associated volume change. Imaging techniques such as in-situ SEM and in-situ electron backscatter diffraction (EBSD) have traditionally been powerful tools to study the microstructure and phase transformations in various systems~\cite{EBSD,insitu,EBSDScope,EBSDTi}. The present in-situ temperature resolved SEM studies were performed on EVO LS10 Zeiss scanning electron microscope, using a commercial DEBEN heating/cooling stage. The temperature fluctuations (after stabilization) were maintained within 0.5 K through a built-in PID controller. The thermocouple, which was mounted on the sample stage, is expected to have been in good thermal contact with the sample due to the conducting nature of the intermediate components (sample stage, carbon tape and the sample itself).

Dramatic shrinkage in the volume of the sample, which is associated with the $\alpha \rightarrow \beta$ transition, was observed at a temperature T$\sim$313~K (see Fig.~\ref{fig:fig1}(a)). A real-time video recording from a different run can be found in supplementary material. The $\beta \rightarrow \alpha$ transition could not be recorded in real time due to its slow transformation rate. Still images were recorded to monitor the changes in a tin sample maintained at 248~K over a period of $\sim$27~h (see Fig~\ref{fig:fig1}(b)). The SEM images show an increase in volume, cracks and micro-fractures, all of which are clear indicators of the $\beta \rightarrow \alpha$ transition. Further imaging studies using in-situ EBSD may be interesting to explore, since the phase composition can be determined from the collected Kikuchi patterns. Thus, it may be possible to study the movement of the phase boundaries using this technique. However, the sample preparation would be challenging as good quality data generally requires polished samples, while the $\alpha$-Sn samples are difficult to polish due to their low transition temperature.
%%%%%%%%%%%%%%%%%%%%%%%%%%%%%%%FIG1 STARTS%%%%%%%%%%%%%%%%%%%%%%%%%%%%%%%%%%%%%%%%%%
\begin{figure} [H]
\centering
\includegraphics[width=\textwidth]{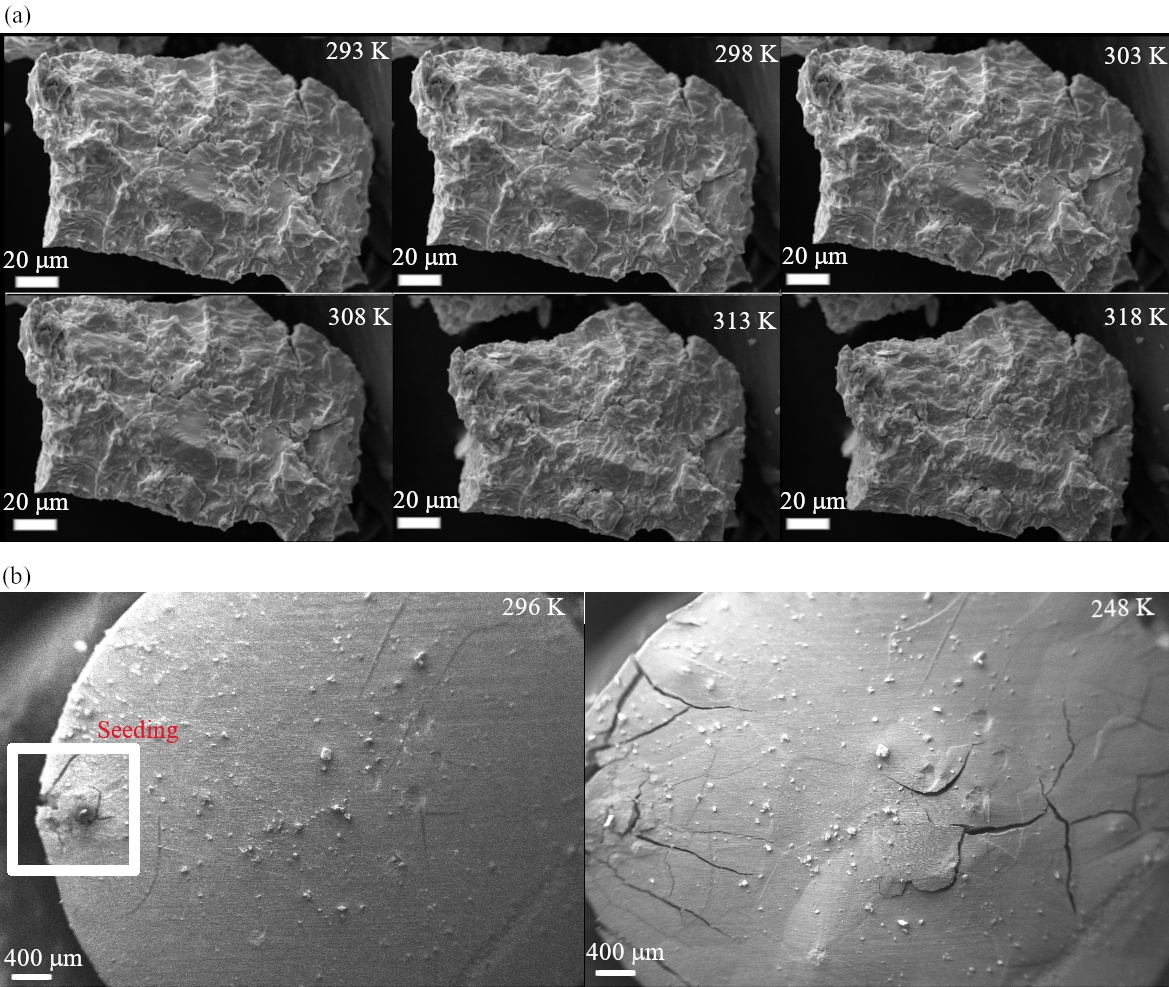}
\caption{\label{fig:fig1}(Colour online) SEM images showing (a) tin sample undergoing the $\alpha \rightarrow \beta$ transition (b) seeded tin sample maintained at 248~K for $\sim$27~h undergoing $\beta \rightarrow \alpha$ transition.}
\end{figure}
%%%%%%%%%%%%%%%%%%%%%%%%%%%%%%FIG1 ENDS%%%%%%%%%%%%%%%%%%%%%%%%%%
The synchrotron XRD measurements were performed at the angle dispersive XRD beamline of the Indus-2, which is a 2.5~GeV synchrotron radiation source at the national facility RRCAT Indore~\cite{RRCAT}. The x-ray beam was tuned to 0.83~$\rm\r{A}$ using a Si(311) double crystal monochromator (DCM). The samples were mounted between kapton tapes and loaded in an in-house developed 2-stage closed cycle refrigerator. The temperature was monitored using a PT100 RTD sensor mounted at the heating element and was controlled using the Lakeshore 331 PID controller. The temperature stability was typically around 0.02 K (for a 60 s exposure) in a step of 2 K. The error in the temperature was $\sim \pm$ 0.5 K, arising mainly from the accuracy and the position of the sensor. The diffraction data were recorded in the transmission mode using the MAR345 detector. It should be mentioned that spinning of the samples was not possible.

For optimizing the use of the available machine time, the heating cycle data (273 - 343~K) were given priority and the cooling cycle data (293 - 213~K) were recorded with a lower density of points. After recording the data at 343~K, the sample was cooled down to 293~K and a final data was recorded after the heating cycle. This was done to check whether the transition reverses on cooling down from 343 to 293~K. Each run included data of the NIST sample $\mathrm{LaB_{6}}$ in the same geometry, for calibration purposes. This data was used to refine the energy of the x-ray used as well as the distance of the sample to the detector. The calibration and conversion of the 2-D image plate diffractograms to 1-D $2\theta$ scans were performed using FIT2D~\citep{1997fit2d, fit2dint}. The 2-D image plate data for the Sn sample can be found in the supplementary material.

\begin{figure} [H]
\centering
\includegraphics[width=\textwidth]{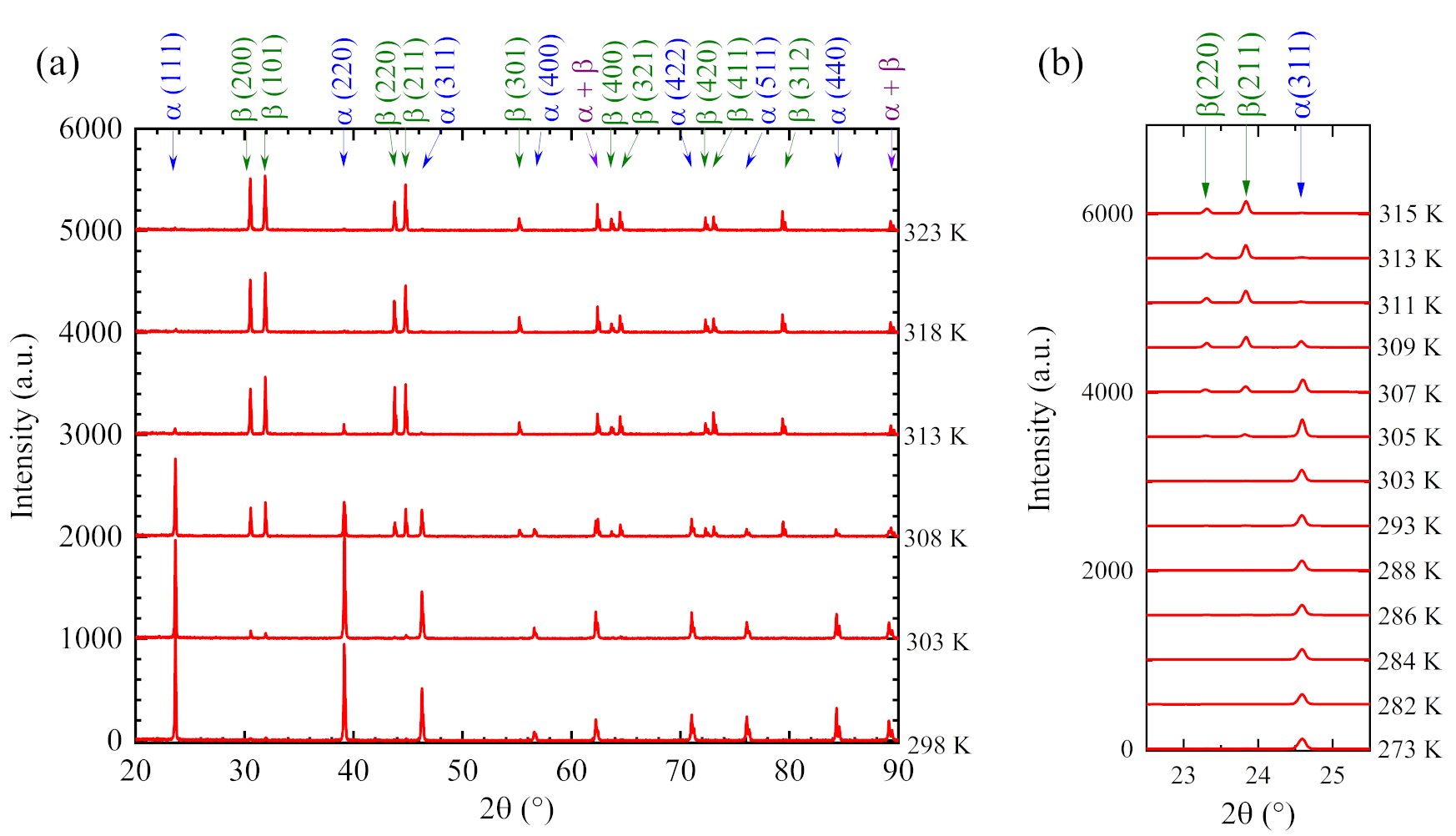}
\caption{\label{fig:fig2}(Colour online) XRD data for the Sn sample showing the $\alpha \rightarrow \beta$ transition. The baseline of the data at different temperatures is shifted for visibility purposes (a) Conventional XRD acquired on Rigaku diffractometer using Cu-$\mathrm{K_{\alpha}}$ x-ray (b) Synchrotron XRD acquired with $\mathrm{\lambda = 0.83 \r{A}}$. }
\end{figure}

\begin{figure} [H]
\centering
\includegraphics[width=0.6\textwidth]{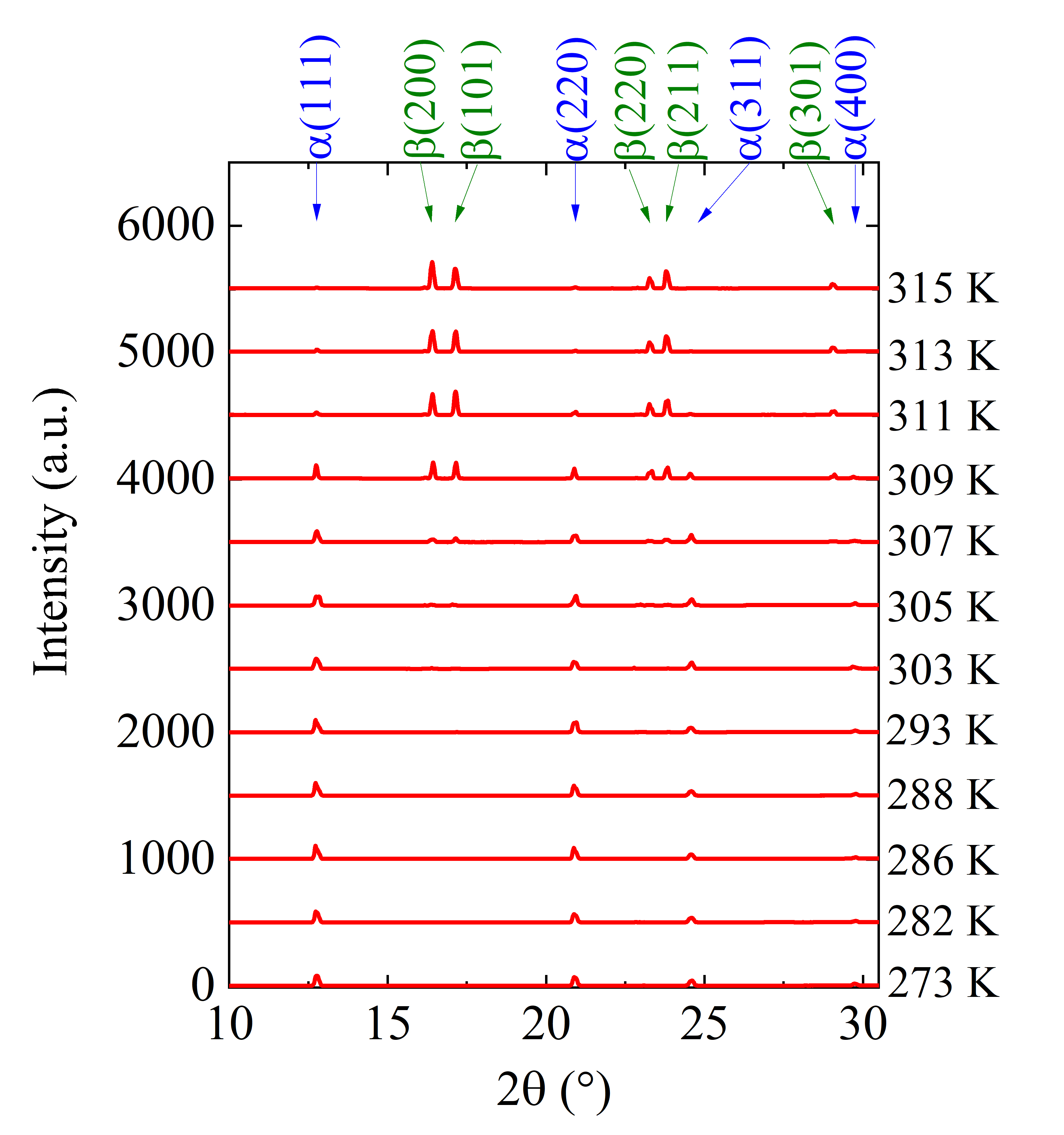}
\caption{\label{fig:fig3}(Colour online) Synchrotron XRD data of the Sn-Cu sample showing the $\alpha \rightarrow \beta$ transition.}
\end{figure}

The disappearance (appearance) of the lines from the $\alpha$-phase ($\beta$-phase) could be tracked in the heating cycle data. Conventional XRD measurements were also performed on a Rigaku SmartLab diffractometer with the SHT1500 heating stage. The heating attachment of the Rigaku diffractometer consists of a platinum sample holder which is surrounded by a furnace. This design is expected to provide a more uniform heating in comparison to a conventional design in which a heating element is in contact with the sample stage. The thermocouple monitoring the sample temperature was in excellent thermal contact through the platinum sample holder. The temperature was controlled through a PID-temperature controller (PTC-EVO) and the data was taken after temperature stabilization. The temperature error for the reported data is within 0.5 K. Both (synchrotron and conventional) XRD data for the Sn sample are shown in Fig.~\ref{fig:fig2}. The synchrotron XRD data for the Sn-Cu sample are shown in Fig.~\ref{fig:fig3}.

The present temperature resolved XRD and synchrotron studies have revealed that the $\alpha \rightarrow \beta$ structural phase transition of Sn and Sn-Cu (0.5\% Cu by weight) occurs between 303~K and 307~K. Alloying with copper at the level of 0.5~\% by weight does not affect the $\alpha \rightarrow \beta$ transition. In case of both the samples, the $\beta$-phase grows slowly from 303~K onwards. The peaks from the $\beta$-phase become distinct around 307~K. With increasing temperature, the peaks from the $\beta$-phase increase in intensity while those from the $\alpha$-phase reduce in intensity. The samples remain in a mixed $\alpha - \beta$ state upto $\sim$315~K, above which the $\alpha$-phase lines disappear. 

\begin{figure} [H]
\centering
\includegraphics[width=\textwidth]{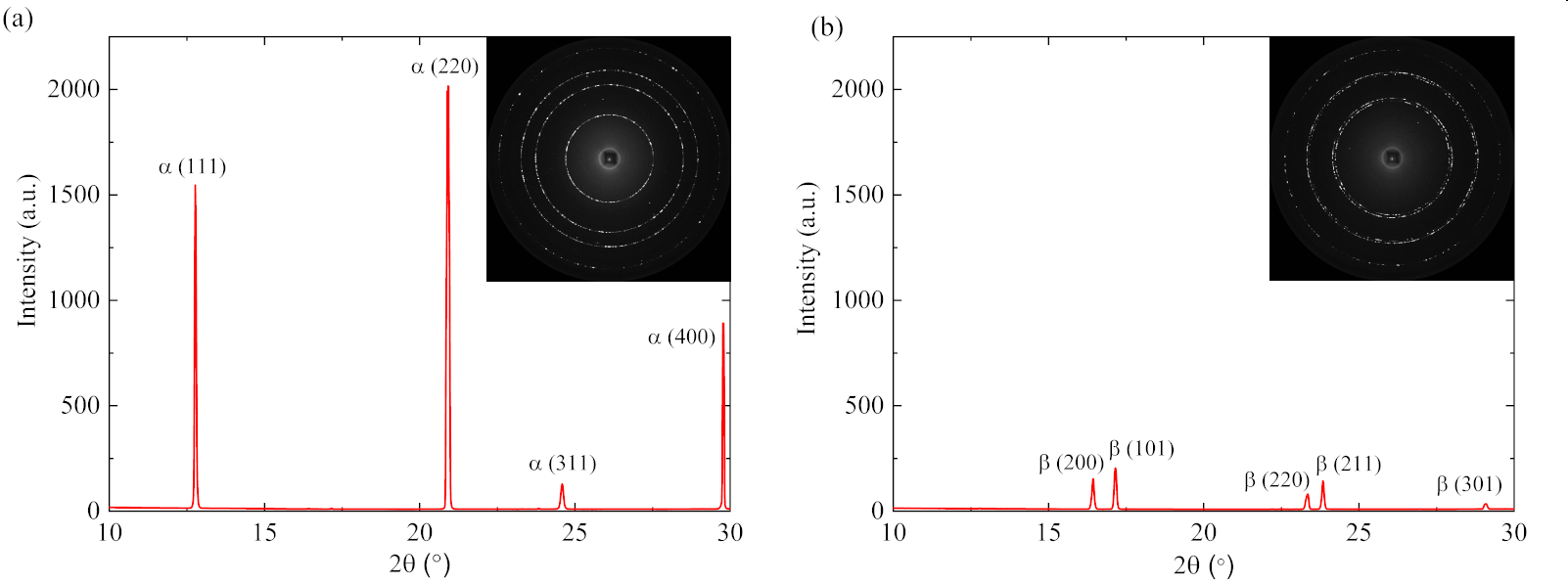}
\caption{\label{fig:fig4}(Colour online) Synchrotron XRD data of the Sn sample at 293~K (a) Before heating from 293 - 343~K (b) After heating to 343~K, the sample was cooled down to 293~K. The reverse $\beta \rightarrow \alpha$ transition was not observed and the lines in (b) were identified to belong to the $\beta$-phase.}
\end{figure}

The transformation does not reverse on cooling from 343~K to 293~K. The comparison between the diffractograms of the Sn sample at 293~K before and after the heating cycle is shown in Fig.~\ref{fig:fig4}. In the previous DSC measurements on these systems, an onset phase transition temperature of 307~K for Sn and 309.3~K for 0.5 \% Sn-Cu (weight \%) was measured ~\cite{TPMRX}. This corresponds to the temperature at which the diffraction peaks from the $\beta$-phase start becoming distinct. The observations from the XRD measurements are consistent with the previous DSC study.
%%%%%%%%%%%%%%%%%%%%%%%%%%%%%%%FIG2 STARTS%%%%%%%%%%%%%%%%%%%%%%%%%%%%%%%%%%%%%%%%%%
%%%%%%%%%%%%%%%%%%%%%%%%%%%%%%FIG2 ENDS%%%%%%%%%%%%%%%%%%%%%%%%%%%%%%%%%%%%%%%%%%%%%
Baking the \textit{TIN.TIN} detector array at a higher temperature like $\sim$323~K for a few min should be sufficient to destroy any $\alpha$-Sn which may have formed during the thermal cycling. However, the risk of a subsequent spontaneous nucleation remains, which is much smaller compared to that of a seeded transformation. This protocol will also be useful for critical circuits that will employ lead-free Sn-Cu solders and operate at low temperatures for long periods. Conversely, $\alpha$-Sn samples should be stored below 303~K in studies where bulk $\alpha$-Sn is the phase of interest.
%%%%%%%%%%%%%%%%%%%%%%%%%%%%%%%FIG3 STARTS%%%%%%%%%%%%%%%%%%%%%%%%%%%%%%%%%%%%%%%%%%
%%%%%%%%%%%%%%%%%%%%%%%%%%%%%%FIG3 ENDS%%%%%%%%%%%%%%%%%%%%%%%%%%%%%%%%%%%%%%%%%%%%%
Although the present study suggests a protocol to minimize the risk of tin pest in a pure tin detector, it may be beneficial for \textit{TIN.TIN} to explore an appropriate tin-rich alloy candidate. It is known that alloying with elements such as Pb, Sb and Bi can stabilize the metallic phase at lower temperatures by pinning lattice defects, thereby inhibiting tin pest by making lattice expansion kinematically unfavourable. Some aspects of the feasibility of using a suitable tin-rich alloy instead of pure tin have been explored in an earlier study~\cite{TPMRX}, in which several tin-rich alloys were prepared and tested for resistance to tin pest. In addition to inhibiting tin pest at a low concentration, a candidate alloy for \textit{TIN.TIN} would have to satisfy some other constraints specific to the experiment. The energy resolution of the detector worsens if the heat capacity of the detector increases. Therefore, it is imperative that the alloy's superconducting transition temperature ($\mathrm{T_{c}}$) should not differ greatly from that of pure tin, as the electronic heat capacity of a superconductor falls exponentially once it is cooled below the $\mathrm{T_{c}}$.  It should be noted that the introduction of an alloying element into the tin matrix may increase the internal background of the detector in the region of interest. This is a critical parameter and the change in background in comparison to pure tin must be critically studied, as the sensitivity of \textit{TIN.TIN} is directly linked to its background. The most promising alloy candidate seems to be Sn-Bi. Under conditions in which pure tin typically develops tin pest within a day, Sn-Bi (0.5\% Bi by weight) has shown no signs of tin pest for $\sim$ 29 months.
%%%%%%%%%%%%%%%%%%%%%%%%%%%%%%%FIG4 STARTS%%%%%%%%%%%%%%%%%%%%%%%%%%%%%%%%%%%%%%%%%%
%%%%%%%%%%%%%%%%%%%%%%%%%%%%%%FIG4 ENDS%%%%%%%%%%%%%%%%%%%%%%%%%%%%%%%%%%%%%%%%%%%%%

In summary, the $\alpha \rightarrow \beta$ structural phase transitions in Sn and Sn-Cu (0.5\% Cu by weight) were investigated using synchrotron XRD. The phase transitions were found to occur between 303 and 307~K, which is significantly higher than the literature value of 286.2~K. Observations from conventional XRD, in-situ SEM and published DSC measurements were consistent with the synchrotron XRD data. Based on these observations, it is expected that baking the \textit{TIN.TIN} detector at $\sim$323~K between thermal cycles will significantly minimize the risk of tin pest.

\section*{Acknowledgments}
We thank the staff at RRCAT Indore for providing beam-time; Dr.~S.~Banik for her help with the transport logistics;  Prof.~S.~Mazumdar, Prof.~R.~Mallik and their students for helping with incubating the samples; Mr.~N.~A.~Kulkarni, Mr.~V.~J.~Mhatre, Mr.~M.~S.~Pose and Mr.~K.~V.~Divekar for help with measurements at TIFR Mumbai. This work was supported by the Department of Atomic Energy, Government of India (Project No. 12-R$\&$DTFR-5.02-0300).

\bibliography{mybib}

\begin{thebibliography}{28}
\expandafter\ifx\csname natexlab\endcsname\relax\def\natexlab#1{#1}\fi
\providecommand{\url}[1]{\texttt{#1}}
\providecommand{\href}[2]{#2}
\providecommand{\path}[1]{#1}
\providecommand{\DOIprefix}{doi:}
\providecommand{\ArXivprefix}{arXiv:}
\providecommand{\URLprefix}{URL: }
\providecommand{\Pubmedprefix}{pmid:}
\providecommand{\doi}[1]{\href{http://dx.doi.org/#1}{\path{#1}}}
\providecommand{\Pubmed}[1]{\href{pmid:#1}{\path{#1}}}
\providecommand{\bibinfo}[2]{#2}
\ifx\xfnm\relax \def\xfnm[#1]{\unskip,\space#1}\fi
%Type = Article
\bibitem[{Xu et~al.(2017)Xu, Chan, Chen, Chen, Wang, Dejoie, Wong, Hlevyack,
  Ryu, Kee, Tamura, Chou, Hussain, Mo, and Chiang}]{TDS_PRL}
\bibinfo{author}{C.-Z. Xu}, \bibinfo{author}{Y.-H. Chan},
  \bibinfo{author}{Y.~Chen}, \bibinfo{author}{P.~Chen},
  \bibinfo{author}{X.~Wang}, \bibinfo{author}{C.~Dejoie},
  \bibinfo{author}{M.-H. Wong}, \bibinfo{author}{J.~A. Hlevyack},
  \bibinfo{author}{H.~Ryu}, \bibinfo{author}{H.-Y. Kee},
  \bibinfo{author}{N.~Tamura}, \bibinfo{author}{M.-Y. Chou},
  \bibinfo{author}{Z.~Hussain}, \bibinfo{author}{S.-K. Mo},
  \bibinfo{author}{T.-C. Chiang}, \bibinfo{journal}{Phys. Rev. Lett.}
  \bibinfo{volume}{118} (\bibinfo{year}{2017}) \bibinfo{pages}{146402}.
%Type = Article
\bibitem[{Madarevic et~al.(2020)Madarevic, Thupakula, Lippertz, Claessens, Lin,
  Bana, Gonzalez, Di~Santo, Petaccia, Nair, Pereira, Van~Haesendonck, and
  Van~Bael}]{TDS_APL}
\bibinfo{author}{I.~Madarevic}, \bibinfo{author}{U.~Thupakula},
  \bibinfo{author}{G.~Lippertz}, \bibinfo{author}{N.~Claessens},
  \bibinfo{author}{P.-C. Lin}, \bibinfo{author}{H.~Bana},
  \bibinfo{author}{S.~Gonzalez}, \bibinfo{author}{G.~Di~Santo},
  \bibinfo{author}{L.~Petaccia}, \bibinfo{author}{M.~N. Nair},
  \bibinfo{author}{L.~M. Pereira}, \bibinfo{author}{C.~Van~Haesendonck},
  \bibinfo{author}{M.~J. Van~Bael}, \bibinfo{journal}{A{PL} Materials}
  \bibinfo{volume}{8} (\bibinfo{year}{2020}) \bibinfo{pages}{031114}.
%Type = Article
\bibitem[{Si et~al.(2020)Si, Yao, Jiang, Li, Zhou, Ji, Huang, Li, and
  Niu}]{TDS_ACS}
\bibinfo{author}{N.~Si}, \bibinfo{author}{Q.~Yao}, \bibinfo{author}{Y.~Jiang},
  \bibinfo{author}{H.~Li}, \bibinfo{author}{D.~Zhou}, \bibinfo{author}{Q.~Ji},
  \bibinfo{author}{H.~Huang}, \bibinfo{author}{H.~Li},
  \bibinfo{author}{T.~Niu}, \bibinfo{journal}{The Journal of Physical Chemistry
  Letters} \bibinfo{volume}{11} (\bibinfo{year}{2020})
  \bibinfo{pages}{1317--1329}.
%Type = Article
\bibitem[{Cornelius et~al.(2017)Cornelius, Treivish, Rosenthal, and
  Pecht}]{TPRev}
\bibinfo{author}{B.~Cornelius}, \bibinfo{author}{S.~Treivish},
  \bibinfo{author}{Y.~Rosenthal}, \bibinfo{author}{M.~Pecht},
  \bibinfo{journal}{Microelectronics Reliability} \bibinfo{volume}{79}
  (\bibinfo{year}{2017}) \bibinfo{pages}{175--192}.
%Type = Article
\bibitem[{Burns(2009)}]{TPFailure}
\bibinfo{author}{N.~D. Burns}, \bibinfo{journal}{Journal of failure analysis
  and prevention} \bibinfo{volume}{9} (\bibinfo{year}{2009})
  \bibinfo{pages}{461--465}.
%Type = Article
\bibitem[{Mazumdar et~al.(2019)Mazumdar, Garai, Krishnamoorthy, Gupta, Reza,
  Thamizhavel, Nanal, Pillay, and Shrivastava}]{TPMRX}
\bibinfo{author}{A.~Mazumdar}, \bibinfo{author}{A.~Garai},
  \bibinfo{author}{H.~Krishnamoorthy}, \bibinfo{author}{G.~Gupta},
  \bibinfo{author}{A.~Reza}, \bibinfo{author}{A.~Thamizhavel},
  \bibinfo{author}{V.~Nanal}, \bibinfo{author}{R.~G. Pillay},
  \bibinfo{author}{A.~Shrivastava}, \bibinfo{journal}{Materials Research
  Express} \bibinfo{volume}{6} (\bibinfo{year}{2019}) \bibinfo{pages}{076521}.
%Type = Article
\bibitem[{Plumbridge(2007)}]{TP2007}
\bibinfo{author}{W.~J. Plumbridge}, \bibinfo{journal}{Journal of Materials
  Science: Materials in Electronics} \bibinfo{volume}{18}
  (\bibinfo{year}{2007}) \bibinfo{pages}{307--318}.
%Type = Article
\bibitem[{Plumbridge(2008)}]{TP2008}
\bibinfo{author}{W.~J. Plumbridge}, \bibinfo{journal}{Journal of electronic
  materials} \bibinfo{volume}{37} (\bibinfo{year}{2008})
  \bibinfo{pages}{218--223}.
%Type = Article
\bibitem[{Peng(2009)}]{TP2009}
\bibinfo{author}{W.~Peng}, \bibinfo{journal}{Microelectronics Reliability}
  \bibinfo{volume}{49} (\bibinfo{year}{2009}) \bibinfo{pages}{86--91}.
%Type = Article
\bibitem[{Kariya et~al.(2001)Kariya, Williams, Gagg, and Plumbridge}]{TPSnCu}
\bibinfo{author}{Y.~Kariya}, \bibinfo{author}{N.~Williams},
  \bibinfo{author}{C.~Gagg}, \bibinfo{author}{W.~Plumbridge},
  \bibinfo{journal}{JOM} \bibinfo{volume}{53} (\bibinfo{year}{2001})
  \bibinfo{pages}{39--41}.
%Type = Article
\bibitem[{Burgers and Groen(1957)}]{TPMech}
\bibinfo{author}{W.~G. Burgers}, \bibinfo{author}{L.~J. Groen},
  \bibinfo{journal}{Discussions of the Faraday Society} \bibinfo{volume}{23}
  (\bibinfo{year}{1957}) \bibinfo{pages}{183--195}.
%Type = Article
\bibitem[{Nanal(2014)}]{TINTIN}
\bibinfo{author}{V.~Nanal}, \bibinfo{journal}{EPJ Web of Conferences}
  \bibinfo{volume}{66} (\bibinfo{year}{2014}) \bibinfo{pages}{08005}.
%Type = Article
\bibitem[{Raynor and Smith(1958)}]{TPDilatometry}
\bibinfo{author}{G.~Raynor}, \bibinfo{author}{R.~Smith},
  \bibinfo{journal}{Proc. R. Soc. Lond. A} \bibinfo{volume}{244}
  (\bibinfo{year}{1958}) \bibinfo{pages}{101--109}.
%Type = Inproceedings
\bibitem[{Zuo and Xian(2013)}]{P20}
\bibinfo{author}{X.~Zuo}, \bibinfo{author}{A.~Xian}, in:
  \bibinfo{booktitle}{2013 14th International Conference on Electronic
  Packaging Technology (ICEPT)}, \bibinfo{organization}{IEEE}, pp.
  \bibinfo{pages}{152--155}.
%Type = Article
\bibitem[{Ojima et~al.(1990)Ojima, Matsumoto, and Taneda}]{P12}
\bibinfo{author}{K.~Ojima}, \bibinfo{author}{H.~Matsumoto},
  \bibinfo{author}{Y.~Taneda}, \bibinfo{journal}{Journal of the Less Common
  Metals} \bibinfo{volume}{157} (\bibinfo{year}{1990})
  \bibinfo{pages}{L15--L18}.
%Type = Article
\bibitem[{Zeng et~al.(2014)Zeng, McDonald, Gu, Sweatman, and Nogita}]{P29}
\bibinfo{author}{G.~Zeng}, \bibinfo{author}{S.~D. McDonald},
  \bibinfo{author}{Q.~Gu}, \bibinfo{author}{K.~Sweatman},
  \bibinfo{author}{K.~Nogita}, \bibinfo{journal}{Philosophical magazine
  letters} \bibinfo{volume}{94} (\bibinfo{year}{2014}) \bibinfo{pages}{53--62}.
%Type = Article
\bibitem[{Gialanella et~al.(2009)Gialanella, Deflorian, Girardi, Lonardelli,
  and Rossi}]{P13}
\bibinfo{author}{S.~Gialanella}, \bibinfo{author}{F.~Deflorian},
  \bibinfo{author}{F.~Girardi}, \bibinfo{author}{I.~Lonardelli},
  \bibinfo{author}{S.~Rossi}, \bibinfo{journal}{Journal of alloys and
  compounds} \bibinfo{volume}{474} (\bibinfo{year}{2009})
  \bibinfo{pages}{134--139}.
%Type = Article
\bibitem[{Mehl et~al.(2020)Mehl, Ronquillo, Hicks, Esters, Oses, Friedrich,
  Smolyanyuk, Gossett, Finkenstadt, and Curtarolo}]{P1}
\bibinfo{author}{M.~J. Mehl}, \bibinfo{author}{M.~Ronquillo},
  \bibinfo{author}{D.~Hicks}, \bibinfo{author}{M.~Esters},
  \bibinfo{author}{C.~Oses}, \bibinfo{author}{R.~Friedrich},
  \bibinfo{author}{A.~Smolyanyuk}, \bibinfo{author}{E.~Gossett},
  \bibinfo{author}{D.~Finkenstadt}, \bibinfo{author}{S.~Curtarolo},
  \bibinfo{journal}{arXiv preprint arXiv:2010.07168}  (\bibinfo{year}{2020}).
%Type = Article
\bibitem[{Legrain and Manzhos(2016)}]{P5}
\bibinfo{author}{F.~Legrain}, \bibinfo{author}{S.~Manzhos},
  \bibinfo{journal}{AIP Advances} \bibinfo{volume}{6} (\bibinfo{year}{2016})
  \bibinfo{pages}{045116}.
%Type = Article
\bibitem[{Pavone et~al.(1998)Pavone, Pasquale, and de~Gironcoli}]{P7}
\bibinfo{author}{P.~Pavone}, \bibinfo{author}{S.~Pasquale, Baroni},
  \bibinfo{author}{S.~de~Gironcoli}, \bibinfo{journal}{Physical Review B}
  \bibinfo{volume}{57} (\bibinfo{year}{1998}) \bibinfo{pages}{10421}.
%Type = Article
\bibitem[{Nogita et~al.(2013)Nogita, Gourlay, McDonald, Suenaga, Read, Zeng,
  and Gu}]{P9}
\bibinfo{author}{K.~Nogita}, \bibinfo{author}{C.~M. Gourlay},
  \bibinfo{author}{S.~D. McDonald}, \bibinfo{author}{S.~Suenaga},
  \bibinfo{author}{J.~Read}, \bibinfo{author}{G.~Zeng}, \bibinfo{author}{Q.~F.
  Gu}, \bibinfo{journal}{Philosophical Magazine} \bibinfo{volume}{93}
  (\bibinfo{year}{2013}) \bibinfo{pages}{3627--3647}.
%Type = Article
\bibitem[{Farahani et~al.(2019)Farahani, Zijlstra, Mecozzi, Ocel{\'\i}k,
  De~Hosson, and van~der Zwaag}]{EBSD}
\bibinfo{author}{H.~Farahani}, \bibinfo{author}{G.~Zijlstra},
  \bibinfo{author}{M.~G. Mecozzi}, \bibinfo{author}{V.~Ocel{\'\i}k},
  \bibinfo{author}{J.~T.~M. De~Hosson}, \bibinfo{author}{S.~van~der Zwaag},
  \bibinfo{journal}{Microscopy and Microanalysis} \bibinfo{volume}{25}
  (\bibinfo{year}{2019}) \bibinfo{pages}{639--655}.
%Type = Article
\bibitem[{Fukino and Tsurekawa(2008)}]{insitu}
\bibinfo{author}{T.~Fukino}, \bibinfo{author}{S.~Tsurekawa},
  \bibinfo{journal}{Materials {T}ransactions} \bibinfo{volume}{49}
  (\bibinfo{year}{2008}) \bibinfo{pages}{2770--2775}.
%Type = Article
\bibitem[{Gourgues-Lorenzon(2009)}]{EBSDScope}
\bibinfo{author}{A.-F. Gourgues-Lorenzon}, \bibinfo{journal}{Journal of
  {M}icroscopy} \bibinfo{volume}{233} (\bibinfo{year}{2009})
  \bibinfo{pages}{460--473}.
%Type = Article
\bibitem[{Seward et~al.(2004)Seward, Celotto, Prior, Wheeler, and
  Pond}]{EBSDTi}
\bibinfo{author}{G.~G.~E. Seward}, \bibinfo{author}{S.~Celotto},
  \bibinfo{author}{D.~J. Prior}, \bibinfo{author}{J.~Wheeler},
  \bibinfo{author}{R.~C. Pond}, \bibinfo{journal}{Acta Materialia}
  \bibinfo{volume}{52} (\bibinfo{year}{2004}) \bibinfo{pages}{821--832}.
%Type = Article
\bibitem[{Sinha et~al.(2013)Sinha, Sagdeo, Gupta, Upadhyay, Kumar, Singh,
  Gupta, Kane, Verma, and Deb}]{RRCAT}
\bibinfo{author}{A.~K. Sinha}, \bibinfo{author}{A.~Sagdeo},
  \bibinfo{author}{P.~Gupta}, \bibinfo{author}{A.~Upadhyay},
  \bibinfo{author}{A.~Kumar}, \bibinfo{author}{M.~N. Singh},
  \bibinfo{author}{R.~K. Gupta}, \bibinfo{author}{S.~R. Kane},
  \bibinfo{author}{A.~Verma}, \bibinfo{author}{S.~K. Deb},
  \bibinfo{journal}{Journal of Physics: Conference Series}
  \bibinfo{volume}{425} (\bibinfo{year}{2013}) \bibinfo{pages}{072017}.
%Type = Article
\bibitem[{Hammersley et~al.(1997)}]{1997fit2d}
\bibinfo{author}{A.~Hammersley}, et~al., \bibinfo{journal}{European Synchrotron
  Radiation Facility Internal Report ESRF97HA02T} \bibinfo{volume}{68}
  (\bibinfo{year}{1997}) \bibinfo{pages}{58}.
%Type = Article
\bibitem[{Hammersley et~al.(1996)Hammersley, Svensson, Hanfland, Fitch, and
  Hausermann}]{fit2dint}
\bibinfo{author}{A.~Hammersley}, \bibinfo{author}{S.~Svensson},
  \bibinfo{author}{M.~Hanfland}, \bibinfo{author}{A.~Fitch},
  \bibinfo{author}{D.~Hausermann}, \bibinfo{journal}{International Journal of
  High Pressure Research} \bibinfo{volume}{14} (\bibinfo{year}{1996})
  \bibinfo{pages}{235--248}.

\end{thebibliography}

\end{document}